\documentclass[twoside]{ilcws07}
\usepackage[latin1]{inputenc}
\usepackage[dvips]{graphicx,epsfig,color}
\usepackage{wrapfig,rotating}
\usepackage{amssymb,amsmath,array}

\begin{document}
\title{
ILC Sensitivity on Generic New Physics in Quartic Gauge Couplings} 
\author{J\"urgen Reuter$^1$
\vspace{.3cm}\\
1- Albert-Ludwigs-Universit\"at Freiburg -  Physikalisches Institut \\
Hermann-Herder-Str.~3, D-79104 Freiburg - Germany
}

\maketitle

\begin{abstract}
  We investigate the potential of the ILC for measuring anomalous
  quartic gauge couplings, both in production of three electroweak
  gauge bosons as well as in vector boson scattering. Any new physics
  that could possibly couple to the electroweak gauge bosons is
  classified according to its spin and isospin quantum numbers and
  parameterized in terms of resonance parameters like masses, widths,
  magnetic moment form factors etc. By a maximum log-likelihood fit,
  the discovery reach of a 1 TeV ILC for scalar, vector and tensor
  resonances is examined.
\end{abstract} 
 
\newcommand{\jrtrace}[1]{\mathop{\rm tr}\left\{#1\right\}}

\section{Parameterization of new physics in terms of resonances}

The Standard Model (SM) with all yet discovered particles (fermions and
gauge bosons) can be described by a non-linear sigma model for the
electroweak (EW) interactions, dictated by the invariance under $SU(2)_L
\times U(1)$ transformations~(see e.g.~\cite{wk,overview}. In this EW
chiral Lagrangian the Higgs boson is absent, and the model has to be 
renormalized order by order, adding new higher-dimensional
operators. Any new physics beyond the SM can then be parameterized in
terms of these operators in a quite generic way. The building blocks
of this (bottom-up) approach are the SM fermions, $\psi$, the
$SU(2)_L$ gauge bosons, $W_\mu^a$, the hypercharge gauge boson,
$B_\mu$, and the nonlinear representation of the Goldstone bosons:
$\Sigma = \exp \left[ \frac{- i}{v} w^a \tau^a \right]$. The
longitudinal vector bosons are built from the Goldstone bosons within
the vector $\mathbf{V} = \Sigma (\mathbf{D} \Sigma)^\dagger$. To describe
isospin-breaking effects, one singles out the neutral component: $\mathbf{T}
= \Sigma \tau^3 \Sigma^\dagger$. With these prerequisites we can write
the minimal SM Lagrangian (without the yet unobserved Higgs boson)
including all the EW gauge interactions as  
\begin{equation*}
  \mathcal{L}_{\text{min}} = \sum_\psi \overline{\psi} (i \gamma^\mu D_\mu)
  \psi - \frac{1}{2 g^2} \jrtrace{\mathbf{W}_{\mu\nu} \mathbf{W}^{\mu\nu}} - 
  \frac{1}{2 g^{\prime 2}} \jrtrace{\mathbf{B}_{\mu\nu} \mathbf{B}^{\mu\nu}} +
  \frac{v^2}{4} \jrtrace{(vD_\mu \Sigma) (vD^\mu \Sigma)}
\end{equation*}
The complete Lagrangian, since non-renormalizable, contains infinitely
many higher-dimen\-sio\-nal operators and, hence, infinitely many
parameters: 
\begin{equation*}
  \mathcal{L}_{\rm eff} = \mathcal{L}_{\text{min}} - \sum_\psi
  \overline{\psi}_L 
  \Sigma M \psi_R + \beta_1 \mathcal{L}'_0 + \sum_i \alpha_i \mathcal{L}_i + 
  \frac{1}{v} \sum_i \alpha_i^{(5)} \mathcal{L}^{(5)} +  
  \frac{1}{v^2} \sum_i \alpha_i^{(6)} \mathcal{L}^{(6)} + \ldots 
\end{equation*}

All of flavor physics is contained in the fermion mass matrix $M$, but
is ignored for the rest of the paper, since we are interested
mainly in the bosonic EW structure. Indirect information on
new physics is encoded in the $\rho$ (or $T$) parameter $\beta_1$, the
$\alpha$ parameters and higher-dimensional coefficients. The
parameters above can be expressed in terms of the fundamental building
blocks (for more details cf.~\cite{resonances}):

\begin{xalignat*}{2}
  \mathcal{L}'_0 &=\; \frac{v^2}{4} \jrtrace{\mathbf{T} \mathbf{V}_\mu} \jrtrace{\mathbf{T} \mathbf{V}^\mu} & &
  \\
  \mathcal{L}_1 &=\; \jrtrace{\mathbf{B}_{\mu\nu} \mathbf{W}^{\mu\nu}}
  &
  \mathcal{L}_6 &=\; \jrtrace{\mathbf{V}_\mu \mathbf{V}_\nu} \jrtrace{\mathbf{T} \mathbf{V}^\mu} \jrtrace{\mathbf{T}
  \mathbf{V}^\nu}
  \\
  \mathcal{L}_2 &=\; i \jrtrace{\mathbf{B}_{\mu\nu} \lbrack \mathbf{V}^\mu , \mathbf{V}^\nu
    \rbrack}
  &
  \mathcal{L}_7 &=\; \jrtrace{\mathbf{V}_\mu \mathbf{V}^\mu} \jrtrace{\mathbf{T} \mathbf{V}_\nu} \jrtrace{\mathbf{T}
  \mathbf{V}^\nu}
  \\
  \mathcal{L}_3 &=\; i \jrtrace{\mathbf{W}_{\mu\nu} \lbrack \mathbf{V}^\mu , \mathbf{V}^\nu
  \rbrack}
  &
  \mathcal{L}_8 &=\; \tfrac14 \jrtrace{\mathbf{T} \mathbf{W}_{\mu\nu}} \jrtrace{\mathbf{T} \mathbf{W}^{\mu\nu}}
  \\
  \mathcal{L}_4 &=\; \jrtrace{\mathbf{V}_\mu \mathbf{V}_\nu} \jrtrace{\mathbf{V}^\mu \mathbf{V}^\nu}
  &
  \mathcal{L}_9 &=\; \tfrac{i}{2} \jrtrace{\mathbf{T} \mathbf{W}_{\mu\nu}} \jrtrace{\mathbf{T} \lbrack
  \mathbf{V}^\mu , \mathbf{V}^\nu \rbrack}
  \\
  \mathcal{L}_5 &=\; \jrtrace{\mathbf{V}_\mu \mathbf{V}^\mu} \jrtrace{\mathbf{V}_\nu \mathbf{V}^\nu}
  &
  \mathcal{L}_{10} &=\; \tfrac12 \left( \jrtrace{\mathbf{T} \mathbf{V}_\mu} \jrtrace{\mathbf{T} \mathbf{V}^\mu}
  \right)^2
\end{xalignat*}

\begin{figure}
  \begin{center}
    \includegraphics[width=.42\textwidth,height=.36\textwidth]
		    {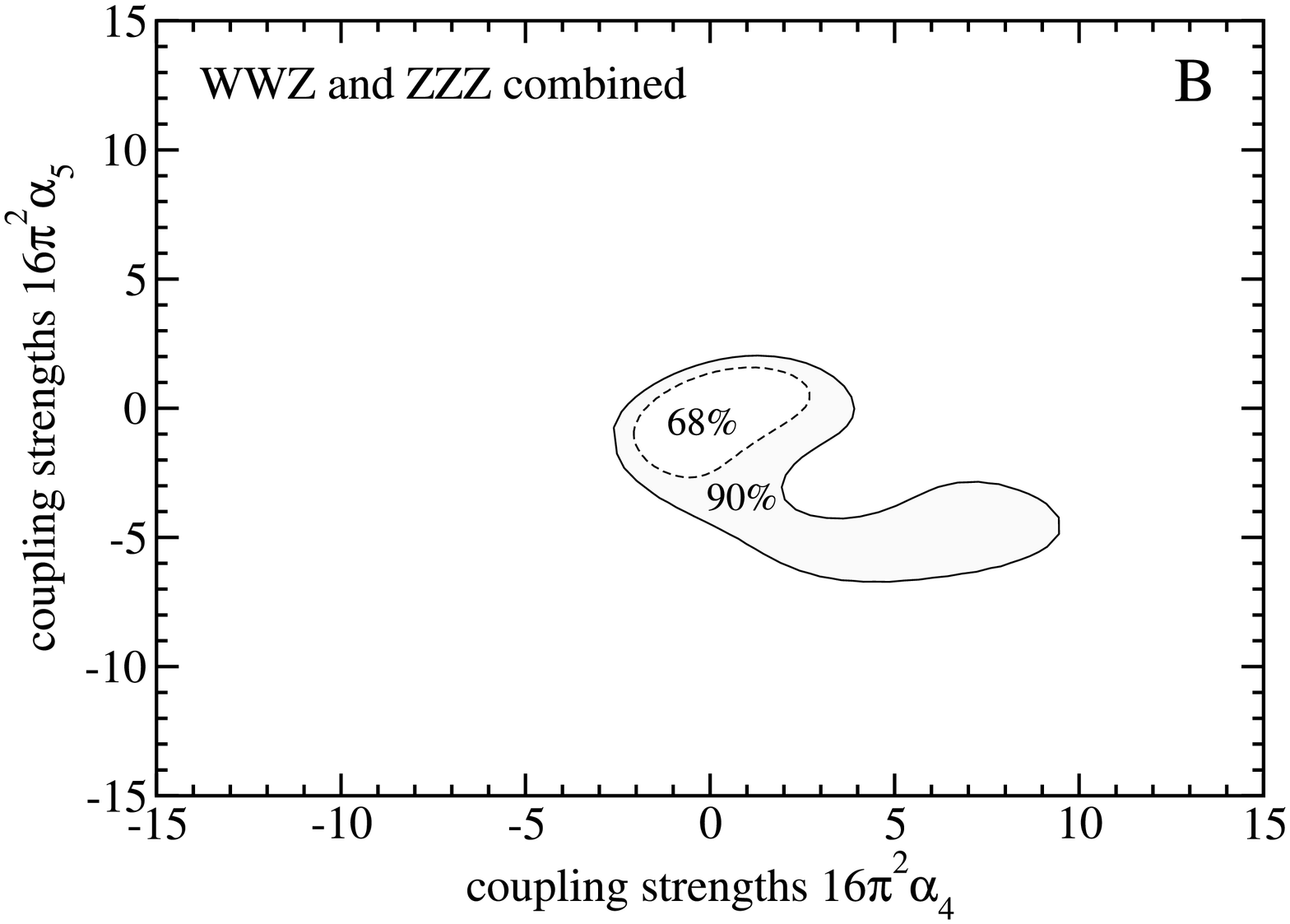}
		    \quad
    \includegraphics[width=.42\textwidth,height=.36\textwidth]
		    {reuter.juergen.qgc.fig2.eps}
  \end{center}
\caption{Left: Combined fit for $WWZ/ZZZ$ production at $\sqrt{s}=$
  1 TeV, 1 ab$^{-1}$, both beams polarized. Right: Expected
  sensitivity (combined fit for all processes) to quartic
  anomalous couplings for a 1 ab$\sp{-1}$ $e^+e^-$ sample in the
  conserved  $SU(2)_c$ case. Solid lines represent $90\%$ CL, dashed
  ones 68\%.}   
\label{fig:triboson_total}
\end{figure}

\begin{table}
  \begin{center}
    \begin{tabular}{|r|ccc|}\hline
      & 
      $J = 0$ & 
      $J = 1$ & 
      $J = 2$ 
      \\\hline
      $I = 0$ & 
      $\sigma^0 \;\text{(Higgs ?)}$ &
      $\omega^0 \; (\gamma'/Z' \; ?)$ &
      $f^0 \; \text{(Graviton ?)}$ 
      \\          
      $I = 1$ & 
      $\pi^\pm, \pi^0 \; \text{(2HDM ?)}$ & 
      $\rho^\pm, \rho^0 \; (W'/Z' \; ?)$ & 
      $a^\pm, a^0$ 
      \\ 
      $I = 2$ & $\phi^{\pm\pm}, \phi^\pm, \phi^0 \;
      \text{(Higgs triplet ?)}$ & 
      $\text{---}$ &
      $t^{\pm\pm}, t^\pm, t^0$ 
      \\\hline
    \end{tabular}
  \end{center}  
  \caption{Classification of resonances that could possibly couple to
    the sector of EW bosons according to their spin and
    isospin quantum numbers, together with some simple examples for them.
    }
  \label{tab:classification}
\end{table}

The $\alpha$ parameters can be measured at ILC with an expected
accuracy at least an order of magnitude better than at LEP, which
allows to access new physics scales that lie outside the kinematical
range of LHC. One of the tasks of this paper is to study the
sensitivity of ILC for new physics scales in the bosonic EW
sector, parameterized by the $\alpha_i$. From the LEP experiments we
already know that the $\alpha$ parameters must be quite small,
$\alpha_i \ll 1$. If new physics coupled to the EW sector is
present, we expect the parameters to be of the order of $\alpha_i
\gtrsim 1/16\pi^2 \approx 0.006$, because the higher-dimensional
operators renormalize divergences which appear with $\mathcal{O}(1)$
coefficients, $16\pi^2 \alpha_i \gtrsim 1$.    

A single new physics scale, $\Lambda$ or $\Lambda^*$, by which the
higher-dimensional operators are suppressed in the form of $\alpha_i
\sim v^2/\Lambda^2$, is in itself not a very meaningful
quantity. Furthermore, it cannot be unambiguously extracted, since the
operator normalization is arbitrary as long as the full theory is
unknown. And, as we will demonstrate below, the power counting can be
quite intricate, such that there is no simple one-to-one
correspondence between new physics and chiral Lagrangian parameters.
  
To be specific: we consider resonances that couple to the EW
symmetry breaking sector of the SM. The resonance masses will give
detectable shifts in the $\alpha_i$ parameters. These resonances could
either be quite narrow in which case we would call them ``particles''
or rather wide where they would be accounted for as a
``continuum''. In that sense, the approach we are using here accounts
for both weakly and strongly interacting models. In
Tab.~\ref{tab:classification} we classified all possibilities of
resonances that can couple to the EW sector according to their spin
and isospin quantum numbers. A special case is the parameter $\beta_1$
(``$\rho$'' parameter) being much smaller than the others as it
expresses the $SU(2)_c$ custodial symmetry almost respected by the SM
Lagrangian. The custodial symmetry is broken by the hypercharge gauge
interactions $g' \neq 0$ and the fermion masses.  

\begin{table}
  \begin{center}
    \begin{small}
    \begin{tabular}{|l|l|r|}      
      \hline 
      $e^+e^-\to$ & Subproc. & $\sigma\;[\text{fb}]\quad$
      \\\hline\hline
      $\nu_e\bar \nu_e q \bar q q \bar q$ &
      $WW \to WW$ &
      23.19\phantom{0}
      \\
      $\nu_e \bar\nu_e q\bar qq\bar q$ &
      $WW \to ZZ$ &
      7.624 
      \\\hline 
      $\nu \bar\nu q \bar qq\bar q$ &
              $V \to VVV$ &
      9.344 
      \\\hline 
      $\nu e q\bar qq\bar q$ &
      $WZ \to WZ$ &
      132.3\phantom{0} 
      \\
      $eeq\bar q q \bar q$ &
      $ZZ \to ZZ$ &
      2.09\phantom{0}   
      \\
      $eeq \bar q q \bar q$ &
      $ZZ \to WW$ &
      414.\phantom{0}\phantom{0}\phantom{0}  
      \\\hline 
      $b\bar b X$ &
      $e^+ e^- \to t \bar t$ & 
      331.768  
      \\\hline 
      $q\bar q q\bar q$ &
      $e^+ e^- \to WW$ &
      3560.108  
      \\
      $q \bar q q \bar q$ &
      $e^+ e^- \to ZZ$ & 
      173.221  
      \\\hline 
      $e\nu q \bar q$ &
      $e^+ e^- \to e\nu W$ & 
      279.588  
      \\
      $e^+ e^- q \bar q$ &
      $e^+ e^- \to eeZ$ & 
      134.935  
      \\\hline
      $X$ &
      $e^+ e^- \to q\bar q$ &
      1637.405  \\ 
      \hline
    \end{tabular}    
    \end{small}
    \quad
    \begin{tabular}{|c|c|c|}\hline
      \multicolumn{3}{|c|}{$SU(2)_c$ conserved
      } \\
      \hline \textbf{\ coupl. \ } & \textbf{ $\sigma
        -$} & \textbf{$\sigma + $}\\                    
      \hline \ $\alpha_4$ \ & \ -1.41 \  &
      \ 1.38 \ \\                    
      \hline \ $\alpha_5$ \ & \ -1.16 \  &
      \ 1.09 \  \\                    
      \hline\hline
      \multicolumn{3}{|c|}{$SU(2)_c$ broken
      } \\                   
      \hline \textbf{\ coupl. \ } & \textbf{ $\sigma
        -$} & \textbf{$\sigma + $}\\                    
      \hline \ $\alpha_4$ \ & \ -2.72 \  &
      \ 2.37 \ \\                                              
      \hline \ $\alpha_5$ \ & \ -2.46 \  &
      \ 2.35 \  \\ 
      \hline \ $\alpha_6$ \ & \ -3.93 \  &
      \ 5.53 \ \\       
      \hline \ $\alpha_7$ \ & \ -3.22 \  &
      \ 3.31 \  \\       
      \hline \ $\alpha_{10}$ \ & \ -5.55 \
      & \ 4.55 \ \\                   
      \hline
    \end{tabular}
  \end{center}
  \caption{Left: Generated processes and cross sections for signal and 
    background for $\sqrt{s}=1$ TeV, polarization 80\% left for
    electron and 40\% right for positron beam.  For each process,
    those final-state flavor combinations are included that correspond
    to the indicated signal or background subprocess. Right:
    The expected sensitivity from 1 ab$^{-1}$ $e^+e^-$ sample
    at 1 TeV, asymmetric 1 sigma errors.}
  \label{tab:vecbosscat}
\end{table}

The most reliable way to take the effects of heavy resonances on the
EW Lagrangian into account is to integrate them out in the
path integral by completing the square in the Gaussian integration. 
Considering the leading order effects of resonances on the EW sector,
integrating out a resonance $\Phi$ generates higher-dimensional
current-current interactions:
\begin{equation*}
  \mathcal{L}_\Phi = z \left[ \Phi \left( M_\Phi^2 + D D \right)\Phi + 2
    \Phi J \right] \quad \Rightarrow \quad 
  \mathcal{L}_\Phi^{\rm eff} = - \frac{z}{M^2} JJ + \frac{z}{M^4} J
    (D D) J + \mathcal{O} (M^{-6}) 
\end{equation*}
Here, $D$ is the covariant derivative with respect to $SU(2)_L \times
U(1)$, $J$ is the current of the bosonic sector of the SM and $z$ is a
normalization constant. The simplest example is a scalar singlet
$\sigma$ with Lagrangian $\mathcal{L}_\sigma = -
\frac12 \sigma (M_\sigma^2 + \partial^2) \sigma - \frac{g_\sigma}{2}
v \sigma \jrtrace{\mathbf{V}_\mu \mathbf{V}^\mu} - \frac{h_\sigma}{2}
\jrtrace{\mathbf{T} \mathbf{V}_\mu} 
\jrtrace{\mathbf{T} \mathbf{V}^\mu}$, which leads to an effective
Lagrangian with the 
following anomalous quartic couplings 
\def\fact{\cdot v^2 / (8M_\sigma^2)}
$\alpha_5 = g_\sigma^2 \fact$, $\alpha_7 = 2 g_\sigma h_\sigma \fact$,
$\alpha_{10} = 2 h_\sigma^2 \fact$. A special case of this would be
the SM Higgs with $g_\sigma = 1$ and $h_\sigma = 0$. (Another example
for such states would be the light pseudoscalars present in Little
Higgs models~\cite{pseudoaxions}). 

\begin{figure}
  \begin{center}
    \includegraphics[width=.7\columnwidth,height=.4\columnwidth]{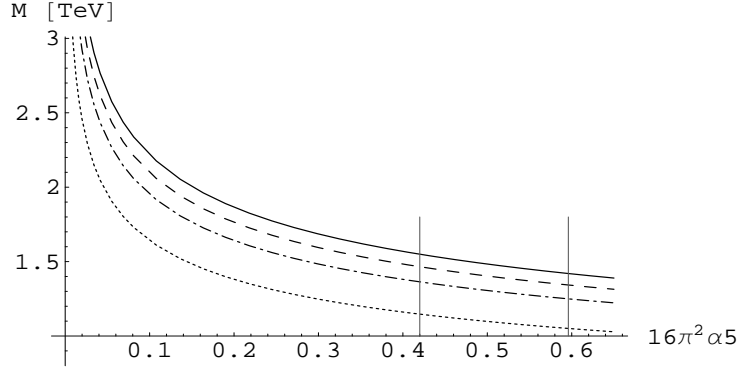}    
  \end{center}
  \caption{
    Mass of the scalar singlet resonance in the
    isospin-conserving case as a function of $\alpha_5$, with the
    resonance's width to mass ratio $f_\sigma$ equal to $1.0$
    as full, $0.8$ as dashed, $0.6$ as dot-dashed, and $0.3$ as
    dotted line, respectively. The left vertical line in the plot
    is the $1\,\sigma$ limit on $\alpha_5$, the right one the
    $2\,\sigma$ limit.}
    \label{fig:su2consscalar}
\end{figure}

Assuming that this scalar resonance is much heavier than the EW gauge
bosons ($M_\sigma \gg M_W, M_Z$) we can neglect mass effects and
calculate its width:
\begin{equation*}
  \Gamma_\sigma = \frac{g_\sigma^2 + \frac12 (g_\sigma^2 + 2
    h_\sigma^2)^2}{16 \pi} \left( \frac{M_\sigma^3}{v^2} \right) + 
  \Gamma (\text{non}-WW,ZZ)
\end{equation*} 
For a broad continuum the largest allowed coupling would result in a
width that equals the resonance's mass, $\Gamma \sim M \gg
\Gamma(\text{non}-WW,ZZ) \sim 0$. This limiting case translates into
bounds for the effective Lagrangian (e.g. in the case of a scalar
singlet with no isospin violation):
\begin{equation*}
  \alpha_5 \leq \frac{4 \pi}{3} \left( \frac{v^4}{M_\sigma^4}
  \right) \approx \frac{0.015}{(M_\sigma \;\text{in TeV})^4}
  \quad \Rightarrow \quad 16 \pi^2 \alpha_5 \leq
  \frac{2.42}{(M_\sigma \;\text{in TeV})^4} 
\end{equation*} 

In performing the power counting in a similar manner for other
resonances one would naively conclude the following dependence of the
anomalous couplings on the resonance masses:
\begin{center}
  \fbox{
    \begin{tabular}{rll}
      Scalar: & $\Gamma \sim g^2 M^3$, $\alpha
      \sim g^2/M^2$ & $\Rightarrow \quad 
      \alpha_{\text{max}}  
      \sim 1 / M^4$   
      \\
      Vector:& $\Gamma \sim g^2 M$, $\alpha
      \sim g^2/M^2$ & $\Rightarrow \quad 
      \alpha_{\text{max}}  
      \sim 1 / M^2$   
      \\
      Tensor:& $\Gamma \sim g^2 M^3$,
      $\alpha  
      \sim g^2/M^2$ & $\Rightarrow \quad 
        \alpha_{\text{max}}  
        \sim 1 / M^4$   
  \end{tabular}}
\end{center}
This naive power counting fails in providing the correct answer (for
the technical details see~\cite{resonances}). Here the $1/M^2$ term
only renormalizes the kinetic energy (i.e. $v$), and hence is  
unobservable. 

So for vector resonances, all $\alpha_i \sim 1/M_\rho^4$, except for
the $\rho$ parameter $\beta_1 \sim \Delta \rho \sim T \sim h_\rho^2 /
M_\rho^2$. Of course, if new physics resonances couple with
non-negligible parameters to the SM fermions, there will be 4-fermion
contact interactions that scale like $j_\mu j^\mu \sim 1/M_\rho^2$ and
constitute effective $T$ and $U$ parameters. Since these are the most
constrained cases (and those most investigated in the literature) we
focus here on physics where these interactions can be neglected
compared to those to the bosonic EW sector. As a remark of caution we 
mention that there is also the possibility of a coupling of the
EW current due to new resonances to the longitudinal EW bosons which
also leads to an effective $S$ parameter $j_\mu V^\mu \sim 1 /
M_\rho^2$. It induces a mismatch between the measured fermionic and
bosonic couplings $g$~\cite{nyff,little_kr}. The presence of heavy
vector resonances leads to the following effects: for the triple gauge
couplings at $\mathcal{O}(1/M^2)$ to a renormalization of the $ZWW$
coupling, at $\mathcal{O}(1/M^4)$ to shifts in $\Delta g_1^Z$, $\Delta
\kappa^\gamma$, $\Delta \kappa^Z$, $\lambda^\gamma$, $\lambda^Z$; for
the quartic gauge couplings at order $\mathcal{O}(1/M^4)$ to shifts in
the $\alpha$ parameters that are orthogonal to the scalar case in the
$\alpha_4$--$\alpha_5$ space.


\section{Results and Interpretation}

There are two ways to study quartic gauge couplings at the ILC, namely
triple boson production and vector boson scattering. Concerning the first
case, we consider the processes $e^+e^- \to WWZ/ZZZ$, which depend on  
the combinations $(\alpha_4 + \alpha_6)$, $(\alpha_5 + \alpha_7)$,
$\alpha_4 + \alpha_5 + 2 (\alpha_6 + \alpha_7 + \alpha_{10})$,
respectively. Polarization populates the longitudinal modes and
drastically suppresses the SM background. The simulations for the
processes discussed here have been performed with the WHIZARD
package~\cite{omega,whizard,omwhiz}, which is ideally suited for 
physics beyond the SM~\cite{omwhiz_bsm}. 

\begin{wraptable}{r}{.5\columnwidth}
  \begin{center}    
    \begin{tabular}{|c||c|c|c|}\hline
      \multicolumn{4}{|c|}{$SU(2)_c$ conserved
      } \\
      \hline
      Spin & $I=0$ & $I=1$ & $I=2$
      \\
      \hline\hline
      $0$ & $1.55$ & $-$ & $1.95$
      \\
      $1$ & $-$ & $2.49$ & $-$
      \\
      $2$ & $3.29$ & $-$ & $4.30$
      \\
      \hline\hline
      \multicolumn{4}{|c|}{$SU(2)_c$ broken
      } \\\hline
      Spin & $I=0$ & $I=1$ & $I=2$
      \\
      \hline\hline
      $0$ & $1.39$ & $1.55$ & $1.95$
      \\
      $1$ & $1.74$ & $2.67$ & $-$
      \\
      $2$ & $3.00$ & $3.01$ & $5.84$
      \\
      \hline
      \end{tabular}
  \end{center}
  \caption{
    Accessible scale $\Lambda$ in TeV for all possible
    spin/isospin channels, derived from the analysis of
    vector-boson scattering at the ILC.} 
  \label{tab:ilcresonancereach}
\end{wraptable}

For the triple boson production we assumed a $1$ TeV ILC with
$1\,\text{ab}^{-1}$ integrated luminosity. The complete six-fermion 
final states generated with WHIZARD have been piped through the
SIMDET fast simulation. As observables we used $M_{WW}^2$, $M_{WZ}^2$, 
and the angle between the incoming electron and the $Z$. We considered
the three cases A) unpolarized, B) 80\% $e^-_R$, C) 80\% $e^-_R$, 60\%
$e^+_L$. One has a branching ratio of 32 \% hadronic decays, for which
we used the Durham jet algorithm. The most severe SM background is
$t\bar{t} \to 6 \;\text{jets}$ being vetoed against by a missing
energy variable cut, $E_\mathrm{mis}^2+p^2_{\perp,\mathrm{mis}}$. So
far, no angular correlations have been used in this analysis yet. The
result is shown for the combined WWZ/ZZZ case in the left of
Fig.~\ref{fig:triboson_total}.    

Vector boson scattering -- as the second process where quartic gauge
couplings could be measured -- has been studied for a 1 TeV ILC with
$1\,\text{ab}^{-1}$, full six-fermion final states, 80 \% $e^-_R$ and
60 \% $e^+_L$ polarization. The contributing channels are mainly
$WW\to WW$, $WW \to ZZ$, $WZ \to WZ$, $ZZ \to ZZ$, in more detail in
the left of Tab.~\ref{tab:vecbosscat}. We performed a binned
log-likelihood analysis for all different spin-/isospin combinations
listed in Tab.~\ref{tab:classification}. To interpret the ILC reach as 
limits on resonances, we consider the width to mass ratio, $f = \Gamma
/ M$, by which we can trade the unknown parameters (i.e.~coupling
constants) by experimentally accessible resonance parameters like the
position and shape of the resonance. 

As the simplest example, we show the $SU(2)$ conserving scalar singlet
in Fig.~\ref{fig:su2consscalar}. Here the relation between the
resonance mass, the $\alpha$ parameters and the width-to-mass ratio, 
$M_\sigma = v \left( (4 \pi f_\sigma) / (3 \alpha_5)
\right)^{\frac14}$, can easily be solved. Extracting limits for
resonances with $SU(2)$ breaking or higher isospin gets more and more
complicated. The most complex case is the $SU(2)$ broken vector
triplet: since the effects from the presence of the vector resonance
enter only at $\mathcal{O}(1/M^4)$ one has to consider all operators
at this order. This includes also magnetic moments of the vector
resonances. Assuming also $SU(2)_c$ breaking the system contains too
many unknown parameters. The missing information can be gained from
the investigation of the triple gauge couplings: we used the
covariance matrix from this measurement~\cite{menges} to find the
minimum in the multi-dimensional parameter space for these cases. 
    
ILC has the ability to detect new physics in the
EW sector even if it is kinematically out of reach. Our
results are summarized in Tab.~\ref{tab:ilcresonancereach}. For the
case of a scalar singlet with conserved $SU(2)$ we combined triple boson
production and boson scattering, shown on the right of
Fig.~\ref{fig:triboson_total} and Tab.~\ref{tab:vecbosscat}. The
limits are translated into resonance masses from the $1\,\sigma$
limits on the $\alpha$s. In general, the limit lies in the range from
$1-6$ TeV, getting better the more internal degrees of freedom are
contributing (higher spin and isospin). It is important to note that
these limits apply for narrow resonances as well as broad continua. 
 
\section{Acknowledgments} 
 
JR was partially supported by the Helmholtz-Gemeinschaft under
Grant No. VH-NG-005.

 
\section{Bibliography} 
 
\begin{footnotesize} 

\end{footnotesize}
 
 
\end{document}